\documentclass[%
 reprint,
 amsmath,amssymb,
 aps,
]{revtex4-1}

\usepackage{verbatim}
\usepackage{filecontents}
\usepackage{graphicx}% Include figure files
\usepackage{dcolumn}% Align table columns on decimal point
\usepackage{bm}% bold math
\newcommand{\be}{\begin{equation}}
\newcommand{\ee}{\end{equation}}
\newcommand{\ba}{\begin{eqnarray}}
\newcommand{\ea}{\end{eqnarray}}
\newcommand\nn{\nonumber}
\newcommand{\vev}[1]{ \left\langle {#1} \right\rangle }

\begin{document}

\title{RNA substructure as a random matrix ensemble}% Force line breaks with \\

\author{Sang Kwan Choi}
  \email{hermit1231@sogang.ac.kr; skchoi@scu.edu.cn}
\affiliation{%
 Center for Theoretical Physics, 
 College of Physical Science and Technology \\
Sichuan University, Chengdu 610064, China
}
\author{Chaiho Rim}%
 \email{rimpine@sogang.ac.kr}
\author{Hwajin Um}%
 \email{um16@sogang.ac.kr}
\affiliation{%
Department of Physics, Sogang University, Seoul 121-742, Korea
}%

\date{\today}% It is always \today, today,

\begin{abstract}
Combinatorial analysis
of a certain abstraction of 
RNA structures has been studied to investigate their statistics.
Our approach regards the backbone of secondary structures
as an alternate sequence of paired and unpaired sets of nucleotides,
which can be described by random matrix model.
We obtain the generating function of the structures 
using Hermitian matrix model  
with Chebyshev polynomial of the second kind
and analyze the statistics
with respect to the number of stems.
To match the experimental findings of the statistical behavior, 
we consider the structures in a grand canonical ensemble 
and find a fugacity value
corresponding to an appropriate number of stems.
\end{abstract}

\maketitle

%\tableofcontents

%===================
\section{Introduction} 
%====================

Ribonucleic acid (RNA) is a single strand of nucleotides,
each of which is one of the four bases, A, U, C and G.
The base pairs are made intra-molecularly,
leading the backbone of nucleotides to form
a 3-dimensional structure called the tertiary structure.
Since the functional role of an RNA is determined
by its tertiary structure,
the prediction of the tertiary
from the sequence of nucleotides
is of great interest \cite{H_2000, FWEG_2017}.

\begin{figure}[!tpb]
\centering
\includegraphics[width=0.48\textwidth]{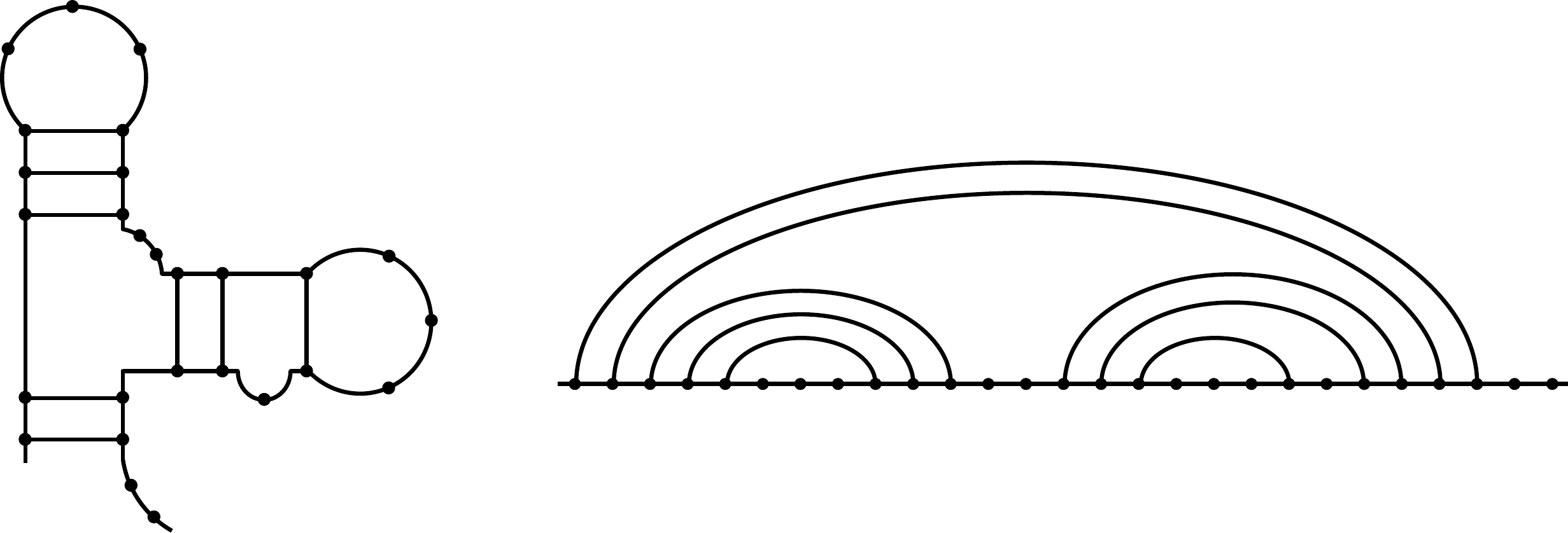}
\caption{\label{fig:second}Secondary structure
and its equivalent graph. 
}
\end{figure}

As an intermediate stage,
the secondary structure 
is a planar structure 
which allows only nested base pairs.
The meaning of the nested base pairs
is evident when the secondary structure
is represented as a labeled graph
over the vertex set $\{1,2,\cdots, n\}$ (Fig.\ref{fig:second}):
the sequence of vertices $(1,2, \cdots, n)$ is put 
on a horizontal line.
The horizontal line represents the backbone and
each vertex denotes the nucleotide.
The base pairs are drawn as arcs in the upper half-plane.
In terms of the graph,
the nested base pairs mean non-crossing arcs.
Another important feature of secondary structures
is that a base pair between adjacent two nucleotides (called {\it1-arc}) 
is not allowed due to the rigidity of the backbone.
In other words, 
any two vertices require at least one unpaired vertex
between them to pair to each other.

Since the secondary interactions are in general
stronger than tertiary interactions such 
as crossing base pairs (called pseudoknots) or base triples,
the prediction of secondary structures
has been intensively studied as a scaffold to the tertiary
\cite{NS_1980, ZS_1981, ZS_1984, RE_1999, SSR_1997}.
The most common approach is the free energy
minimization. The energy of the structures is lower 
as base pairs are formed and it is assumed
that structures tend to be thermodynamically stable.

On the other hand, 
combinatorial approaches 
have also played an important role 
in better understanding the RNA structures.
The structures are often considered
as combinatorial objects such as the graph in Fig.\ref{fig:second},
regardless of the types of bases.
Combinatorial analysis are then applied
to enumerate such objects under various kinds of restrictions
and classifications,
which help to develop and advance the prediction algorithms
\cite{W_1978, HSS_1998, LR_2016}.

In this paper, we investigate a certain substructure
of the secondary structure from the combinatorial point of view,
using the Hermitian random matrix model.
The matrix model was first introduced in RNA structures
to deal with the pesudoknots \cite{VOZ_2005, ACPRS_2012, ACPRS_2013}.
since its topological expansion
facilitates the enumeration of pseudoknot structures
under a topological classification.
The underlying relation between
the matrix model and RNA structures
is established by
a diagrammatic representation of the matrix model.
Although we also employ the matrix model,
we consider here only planar structures
and the diagrammatic representation is mainly used
to describe our substructures.

Before introducing the substructure,
let us first define auxiliary concepts
required to describe the substructure.
An {\it island} is defined as a set of maximally consecutive
paired nucleotides
while a {\it bridge} is a set of maximally consecutive
unpaired nucleotides.
Consequently, the backbone
of the secondary structure
is represented as an alternate sequence of island and bridge.
Then the total number of nucleotides  
$\sum_{i=1}^I ( \ell_i +b_i) $ 
where   $\ell_i$ and $b_i $ are the number of nucleotides 
in the $i$-th island and  bridge, respectively
where we identify the bridge before 
the first island with the one after the last island
for convenience.
We remark here that
the condition of 1-arc forbidden in the secondary structures
implies an important feature of the island,
which is that
there is no base pair 
between vertices inside one island by definition.

\begin{figure}[!tpb]
\centering
\includegraphics[width=0.38\textwidth]{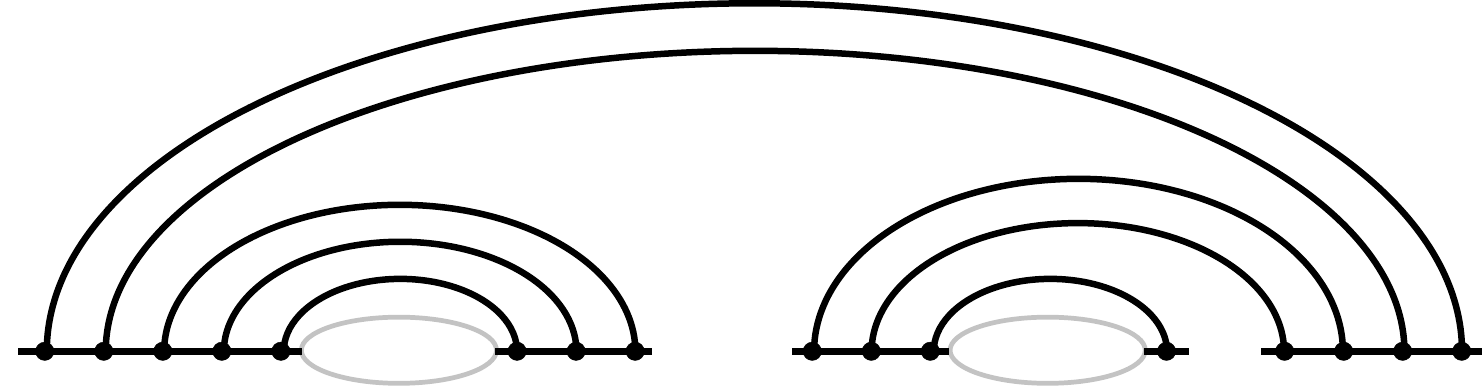}
\caption{\label{fig:island}
The island diagram derived from the secondary structure
given in Fig.\ref{fig:second}.
Each island is depicted as a line segment.
Each bridge is represented as a blank in the island diagram.
A bridge, of which left and right vertices are paired to each other,
is called the hairpin loop.
The two encircled bridges are the hairpin loops.
}
\end{figure}

It is usual to analyze the combinatorics of the secondary structures
with the total number of nucleotides fixed.
However, in this paper,
we are interested in the combinatorics 
of the structures for a given number of base pairs,
regardless of the number of unpaired nucleotides.
Thus, we introduce a substructure ignoring the unpaired ones.
The structure which we call {\it island diagram}
is obtained from the secondary structure
by representing each bridge as a single blank (Fig.\ref{fig:island}).
Accordingly, the island diagram 
retains the configuration of base pairs and unpaired regions,
but is the abstract structure of
the secondary structures with different number of nucleotides in
the unpaired regions.

The paper is organized as follows. In section \ref{sec:theory},
we establish the relation between the matrix model
and the island diagrams. The generating function
enumerating island diagrams is obtained using the matrix model description.
In section \ref{sec:app},
we investigate a distribution of island diagrams,
which disagrees with the one expected from the energy minimization scheme.
A parameter analogous to the chemical potential is introduced
to fit the distribution 
and we give a possible interpretation to the parameter.
Section \ref{sec:pe}
discusses possible extensions of the island diagram configuration
and section \ref{sec:con} is the conclusion.

%그러니까 일반적으로 알엔에이 관련해서 카운팅할때는
%기븐n에서 카운팅 하는 게 일반적이다라는거지
%근데 우리는 loop에 몇 개가 들어있는지는 신경끄고
%loop configuration만 유지하고 있는 다이어그램을 
%기븐 링크 수에 대해 생각해 볼거란 얘기지

%island diagram 하나가 동일한 개수의 sec. str. 를 abstract한다고 가정하고
%계산된 거나 다름없지. practically 말이 안 된다.
%알려진대로 k를 더 늘리는 걸 선호하지 않고
%k가 더 적은 island diagram에 더 많은 sec. str.가 대응될테고
%뮤가 이걸 반영하는 것으로 생각할 수 있다는 것이지. (stack energy외에도)

%k에 따라서 얼마나 더 많은 거를 abstract하는지 quantify할 수 있나?

%===================
\section{Theory} 
\label{sec:theory}
%====================

%%%%%%%%%%%%%%
\subsection{Matrix model description}
\label{HMM}

We employ Hermitian matrix model
to describe and enumerate the island diagrams.
The combinatorial aspect of the matrix model is based on its
diagrammatic representation, which is often called Feynman diagram.
In this section, we present the connection
between the matrix model diagrams and island diagrams.

\begin{figure}[!h]
\centering
\includegraphics[width=0.40\textwidth]{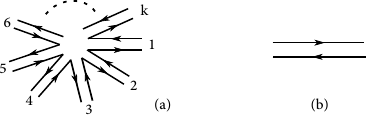}
\caption{\label{fig:vertex}
(a) Vertex of $k$ half-edges. (b) Propagator.}
\end{figure}

Let us first briefly review the diagrams
generated by the matrix model.
The Gaussian expectation value of an operator $O(M)$
is written as
 \be
  \vev{O(M)}
:=\frac1{Z}\int d M  \,   \, O(M) \,
e^{-\frac{N}2 \mathrm{Tr}M^2}
\ee
where $M$ is $N \times N$ Hermitian matrix
and $Z$ is the normalization factor requiring $\vev 1=1$.
It is well-known that the expectation value
can be formulated pictorially.
Let us see this through the example of
$O(M)=N \, \mathrm{Tr} M^k$.
The $\mathrm{Tr} M^k$ corresponds to 
a vertex of $k$ double line half-edges
with a cyclic ordering (Fig.\ref{fig:vertex}(a)) (not to be confused with
the vertex representing a nucleotide
in the secondary structures).
The Gaussian integral indicates pairing the half-edges
with the propagator, represented as the double line edge (Fig.\ref{fig:vertex}(b)).
A diagram obtained by pairing all the half-edges is called the fatgraph
and the expectation value counts the number of fatgraphs.

\begin{figure}[!h]
\centering
\includegraphics[width=0.40\textwidth]{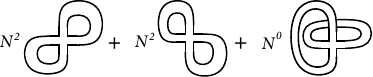}
\caption{\label{fig:M4matrix}
Diagrammatic representation of the expectation value of $M^4$.}
\end{figure}

The advantage of the matrix model 
is its topological expansion,
namely, it counts the number of fatgraphs
filtered with their genus.
The genus $g$ of a fatgraph
is determined from 
its Euler characteristic 
$\chi=2-2g=v-e+f$
where $v$, $e$ and $f$
are the number of vertices, edges 
and faces (loops), respectively.
The genus filtration appears in the expectation value
as the factor $N^{\chi}$
(for rigorous computations and arguments, see for instance \cite{DiF_1999}).
As an explicit example, in the case of $O(M)=N\,\mathrm{Tr} M^4$,
the result is given by
$2 N^2+N^0$ that is pictorially described in Fig.\ref{fig:M4matrix}.
The term $2 N^2$ reads the two planar diagrams
and  $N^0$  reads the non-planar (torus) diagram. 

\begin{figure}[!h]
\centering
\includegraphics[width=0.46\textwidth]{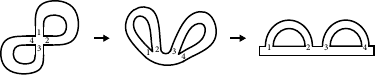}
\caption{\label{fig:transform}
Transformation of a fatgraph into its dual.}
\end{figure}

\begin{figure}[!h]
\centering
\includegraphics[width=0.48\textwidth]{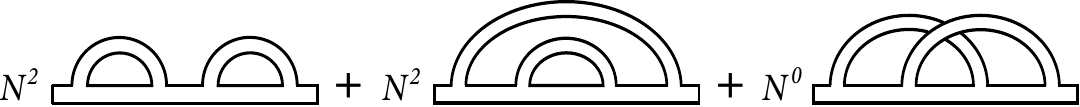}
\caption{\label{fig:M4pic}
Dual representation of the diagrams in Fig.\ref{fig:M4matrix}.}
\end{figure}

In order to see the connection between the fatgraphs
and island diagrams clearly,
it is more convenient 
to use dual graphs of the matrix model diagrams. 
Given a fatgraph of the matrix model,
its dual graph is obtained by
transforming the vertex of $k$ half-edges 
into the horizontal line of $k$ vertices.
In other words, 
the vertex is stretched out 
to form the horizontal line (Fig.\ref{fig:transform}).
In the dual representation, therefore,
the expectation value of $\mathrm{Tr} M^4$
is described as in Fig.\ref{fig:M4pic}.

Let us now find the matrix model description
to represent the island diagrams.
It is obvious that 
we need only genus zero contributions
in the topological expansion of the matrix model
since the secondary structures are the planar structures.
The next step is to find the way
to impose the configuration of islands
on the matrix model diagrams.
Let us consider the number $I$ of islands
each of which has $k_a \geq 1$ vertices
for $a \in \{1,\cdots, I\}$ with $k=k_1+\cdots+k_I$.
We may first try with 
$\big\langle \mathrm{Tr} \prod_{a=1}^I M^{k_a} \big\rangle_0$
where the subscript $0$ denotes genus zero contributions.
Here $M^{k_a}$ corresponds to $k_a$ half-edges
among $k$.
The expectation value is merely rewriting of 
$ \vev{\mathrm{Tr} M^k}_0$
and hence generates all possible
planar diagrams paring $k$ vertices.

\begin{figure}[!h]
\centering
\includegraphics[width=0.40\textwidth]{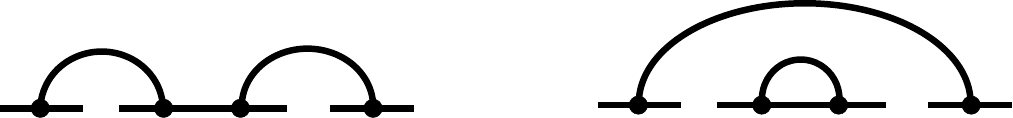}
\caption{\label{fig:M4island}
Graphs described by $M M^2 M$. The graph on the right hand side
is not the island diagram due to the 1-arc in the middle.}
\end{figure}

However, some of the planar diagrams
may not be allowed as island diagrams.
One can see this clearly through
the example of $ \vev{\mathrm{Tr} M M^2 M}_0$,
which generates the first two
planar diagrams in Fig.\ref{fig:M4pic}.
Meanwhile, 
the island configuration we have in mind
by rewriting $M^4$ into $M M^2 M$
is 3 islands with 1, 2 and 1 vertices.
Then, the two planar diagrams
are interpreted as the graphs in Fig.\ref{fig:M4island}.
Recall that, however, secondary structures 
do not allow 1-arc
and hence any base pair is forbidden on one island.
Therefore, the second graph in Fig.\ref{fig:M4island} 
should be excluded from the island diagrams.

In order to impose the constraint,
we introduce $U^{(k_a)} (M)$ instead of $M^{k_a}$
as an operator corresponding to the island with $k_a$ vertices
and consider the expectation value of
\be
\label{op:strand}
O^{(I)}_{\{k_a\}} = 
  \mathrm{Tr} \prod_{a=1}^I U^{(k_a)} (M) \,.
\ee
To find the faithful expression for $U^{(k_a)}$,
we need to impose the condition of 1-arc absent on an island.
This can be done by means of the inclusion-exclusion method,
which gets rid of diagrams with 1-arcs step by step.
The operator $M^{n}$ reduces to 
$M^{n-2}$ after removing 1-arc paring.
The number of ways of assigning  one  1-arc is  $n-1$. 
Note that the expectation value of $M^{n-2}$ 
generates all the diagrams having the 1-arc from $M^n$.
Therefore, we subtract those configurations 
and introduce the operator $M^n-(n-1)M^{n-2}$. 
However, in which case,
the diagrams with two 1-arcs are over-subtracted 
and need to be compensated by adding the term  $M^{n-4}$. 
In this way, adding/subtracting $M^{n-2p}$
with the number of possible contractions of $p$ 1-arcs,
we arrive at the final result,  
\be
U^{(n)}(M)=\sum_{p=0}^{\lfloor n/2 \rfloor}
(-1)^p \binom{n-p}{p} M^{n-2p} \,.
\ee

Finally, we find a simple integration formula 
for the expectation value by identifying 
the polynomial $U^{(n)}$  with  
the Chebyshev polynomial of the second kind $U_n$, 
$U^{(n)}(2\xi)=U_n(\xi)$. 
The  Chebyshev polynomial
has the  product rule,
\be
U_m (\xi) U_n (\xi)=\sum_{k=0}^n  U_{m-n+2k}(\xi) 
~~~~ \mathrm{for} ~~~ n \leq m ,
\ee
and, therefore, the vertex operator 
$ O^{(I)}_{\{k_a\}} $ in \eqref{op:strand}  
ends up with the linear combination of Chebyshev polynomials.
Note that $ \big\langle \mathrm{Tr} U_{n}(M/2) \big\rangle_0 =0$
unless $n=0$ due to the constraint of 1-arc absence. 
Therefore, the expectation value  of  $ O^{(I)}_{\{k_a\}} $ 
is given as the coefficient of $ U_0=1$,
\be
\big\langle  O^{(I)}_{\{k_a\}}\big\rangle_0
=\frac2 \pi \int_{-1}^{1} \prod_{a=1}^{I} U_{k_a}(\xi) U_0(\xi) 
\sqrt{1-\xi^2} d\xi 
\label{vev}
\ee
where the orthonormality of the Chebyshev polynomials is used:
\be 
\int_{-1}^1 U_k(\xi) U_\ell (\xi) \sqrt{1-\xi^2} d\xi
=\frac{ \pi}2 \delta_{k,\ell} .
\ee

%%%%%%%%%%%%%%%%%%%%
\subsection{Generating functions}

Based on the integration formula \eqref{vev}, we can find 
the generating function of the island diagrams.
We first classify the island diagrams
by the number of islands, basepairs,
and hairpin loops. 
A {\it hairpin loop} is defined as 
a bridge, of which veritces on the left and right adjacent islands 
are connected to each other (Fig.\ref{fig:second}).

\begin{figure}[!htpb]
\centering
\includegraphics[width=0.44\textwidth]{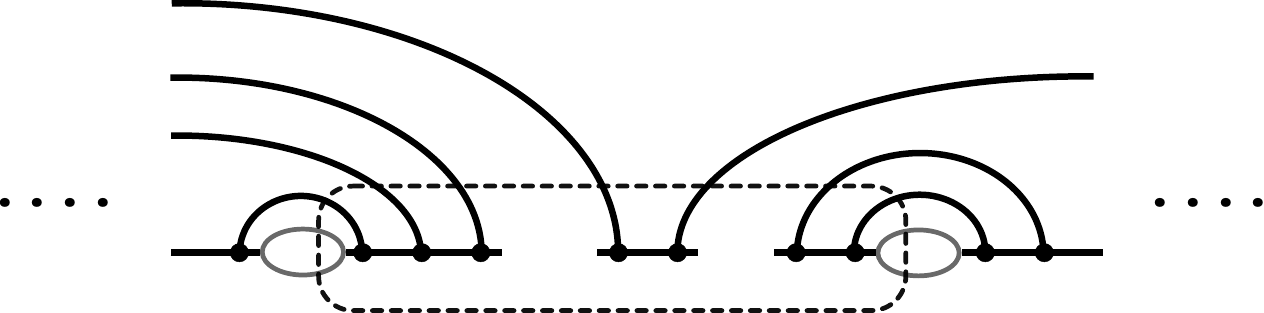}
\caption{\label{fig:effective}
Example of effective island.
The three islands encircled by the dashed line
is one effective island.}
\end{figure}

Let us denote by $ g (h,I,\ell)$
the number of island diagrams
with $h$ hairpins, $I$ islands and $\ell$ basepairs.
It is easy to find $ g (h,I,\ell)$ 
if one uses a property of hairpin loop.  
Observe that 
there must be at least one hairpin loop
between two nucleotides at each end of a basepair.
Therefore, we may regroup islands 
between two consecutive hairpin loops
as one effective island
since there is no basepair connected inside an effective island (Fig.\ref{fig:effective}).
The number of ways to make
an effective island consisting of
$I_a$-islands and $k_a$-nucleotides is given by
$\binom{k_a-1}{I_a-1}$.
Since there are $h+1$ effective islands,
one concludes that 
\begin{align}
g(h,I,\ell)=&\sum_{ \{k_a,  I_a \}}
\prod_{a=1}^{h+1} \binom{k_a-1}{I_a-1}\nn  \\ 
&\times \vev{\mathrm{Tr} \, 
U_{k_1-1}  \prod_{j=2}^{h} U_{k_{j}-2} \, U_{k_{h+1}-1}}_0
\label{g(h,I,ell)}
\end{align}
where
$k_a$ and $l_a$ are 
 constrained as  $k_1+\cdots+k_{h+1}= 2\ell$ 
and $I_1+\cdots+I_{h+1}=I$.
This complicated expression 
is simplified 
if one uses the generating
function of the Chebyshev polynomial,
\be
\sum_{k\ge 0} \sum_{i=0}^k \binom{k}{i} \, z^{k/2}\, y^i \, U_k(\xi)
=\frac1{1-2 \sqrt{z}(1+y)\xi+z(1+y)^2} \,.
\ee 
Then the integral expression in \eqref{g(h,I,ell)} 
 is put into a compact form of 
 the generating function,
\begin{align}
&G(x,y,z):=\sum_{h,I,\ell}  g(h,I,\ell) x^h  y^I  z^\ell \\
& =\sum_h \frac2 \pi \int_{-1}^{1} 
\frac{\sqrt{1-\xi^2}}{\left(1-2 \sqrt{z}(1+y)\xi+z(1+y)^2\right)^{h+1}}
 d\xi  \nn \,.
\end{align}
One can calculate the integral for given $h$ and obtain 
\begin{align}
G(x,y,z)=&\sum_{h} x^h \,z^h\, y^{h+1}\,(1+y)^{h-1} \nn \\ 
&~~~\times{}_2F_1(h+1,h;2;z(1+y)^2).
\label{GXYZ}
\end{align}
Here ${}_2F_1(a,b;c;z)$ is the hypergeometric function
and can be written as ${}_2F_1(h+1,h;2;z)=\sum_{k \ge 0} N(h+k,h)z^k$
where $N(a,b)=\frac1{a}\binom{a}{b}\binom{a}{b-1}$ is the Narayana number
whose generating function is known (see for instance \cite{BH_2011}).
Therefore, one can rewrite $G(x,y,z)$
in the closed form
\begin{align}
&G(x,y,z)=\left( \frac{y}{1+y} \right) \\
&~~\times\frac{1-A(1+B)-\sqrt{1-2A(1+B)+A^2 (1-B)^2}}{2A}  \nn
\end{align}
where $A=z(1+y)^2$ and $B={x \,y}/(1+y)$.

The power series of the generating function
provides the information on $ g (h,I,\ell)$
which will be the building block of other generating functions.
For example, to investigate further details of the secondary structures,
one may introduce the concept of {\it stem} (or {\it stack})
defined as a set of maximally consecutive parallel basepairs.
The {\it length} of a stem is the number of basepairs in the stem.
Let us add one more variable, the number  $k$ of stems  to $ g (h,I,\ell)$
so that the number of configurations 
together with stems are denoted by  $f(k,h,I,\ell)$.
Finding  $f(k,h,I,\ell)$ is based on the number of single-stack diagrams,
where we define the single-stack diagram
as the island diagrams that consists
of only stems of stack-length one.
Namely, each stem is itself a basepair in the single-stack diagrams.
Let $s(h,I,k)$ denotes
the number of the single-stack diagrams of $k$ stems.
The structures of $k$ stems and $\ell$ basepairs
can be constructed
by stacking $\ell-k$ basepairs 
on each stem of the single-stack diagrams.
Therefore, we have the   relation
\be
f(k,h,I,\ell)= \binom{\ell-1}{k-1}  \, s(h,I,k) \,.
\ee
This shows that the generating function 
$F(u,x,y,z):=\sum_{k, h,I,\ell} f(k,h,I,\ell) \, u^k x^h y^I z^{\ell}$ 
is given as 
\be 
F(u,x,y,z)=S\Big(x,y,\frac{u \, z}{1-z}\Big) \,.
\ee
where 
$S(x,y,z):=\sum_{h,I,\ell} s(h,I,\ell) \, x^h y^I z^{\ell}$.
Finally, noting  that $F(1,x,y,z)=G(x,y,z)$, 
one finds  $S(x,y,z)=G(x,y,{z}/(1+z))$ and therefore,
\be
F(u,x,y,z)=G\Big(x,y,\frac{u\, z}{1+ u\, z -z}\Big) \,.
\label{FUXYZ}
\ee

%===================
\section{Application} 
\label{sec:app}
%====================

\begin{figure}[!tpb]
\includegraphics[width=0.48\textwidth]{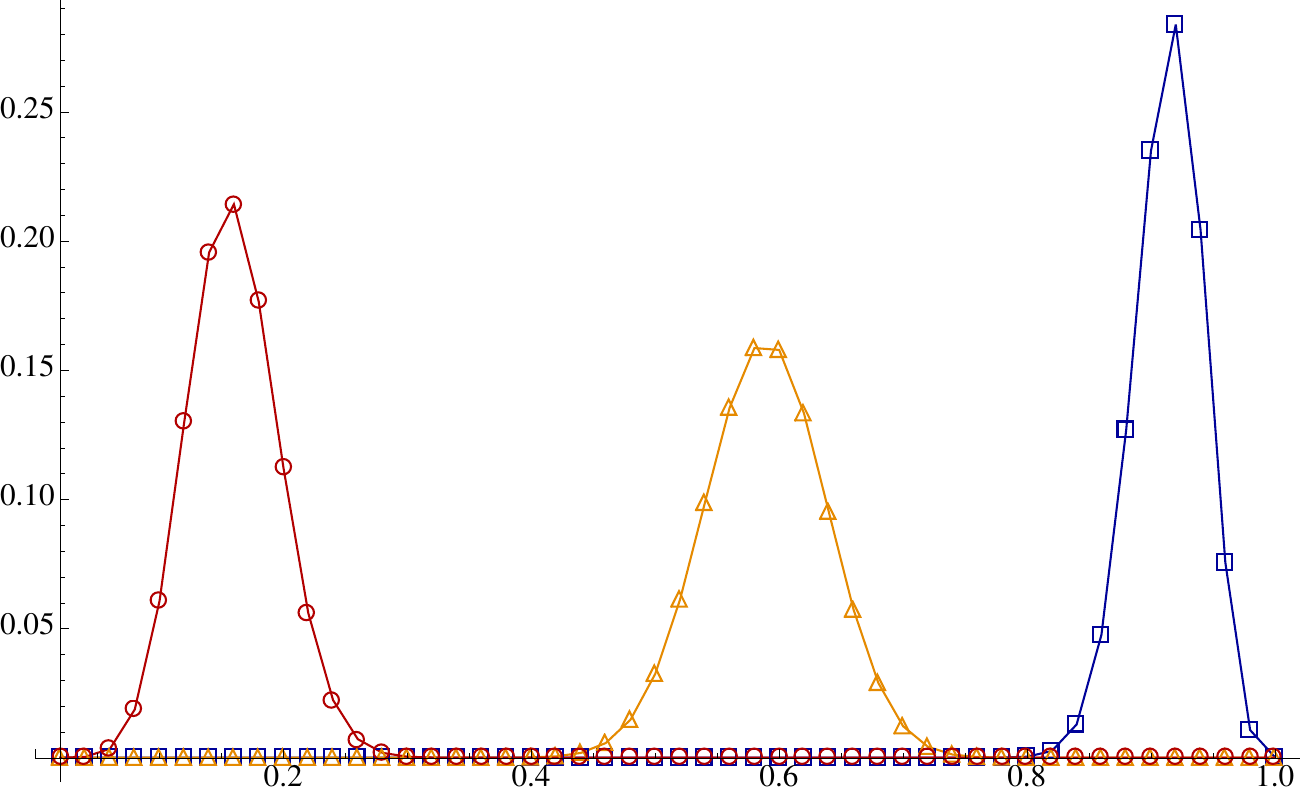}
\centering
\caption{\label{fig:omega} 
Normalized distribution $\Omega_{k,\ell}\, e^{\mu \, k}
/(\sum_k \Omega_{k,\ell}\, e^{\mu \, k})$ of island diagrams
as a function of $k/\ell$ for $\ell=100$:
Square($\mu=0$), 
triangle($\mu=-2$), circle($\mu=-4$).}
\end{figure}

Let us investigate the result of the generating functions.
We consider here the stem distribution of island diagrams for given 
numbers of basepairs.
Using $f(k,h,I,\ell)$, one may define $\Omega_{k,\ell}=\sum_{h,I} f(k,h,I,\ell)$
which represents  the number of island diagrams with $k$ stems 
and $\ell$ basepairs.
From \eqref{GXYZ} and \eqref{FUXYZ}, one finds the explicit formula
\be
\Omega_{k,\ell}=\binom{\ell-1}{k-1}
\sum_{p=0}^{\lfloor \frac{k-1}2 \rfloor}
M(k-1,p) ~ 2^{3p}\, 5^{k-2p-1}
\ee
where
$M(\alpha,\beta):=
\frac{\alpha !}{(\alpha-2\beta)!\, \beta!\, (\beta+1)!}$ 
is the Motzkin polynomial coefficient \cite{M_1948, DS_1977}.
For a given value of $\ell$, one can plot  $\Omega_{k,\ell}$
as a function of $k/\ell$
to see the stem distribution
(square marks in Fig.\ref{fig:omega} for $\ell=100$).
The most probable value appears near $k/\ell=1$.

On the contrary, according to the experimental findings, 
the average basepair per stem in the secondary structures 
is in general greater than two, that is, $k/\ell < 1/2$.
The noticeable difference in the average value of $k/\ell$
tells that the stem distribution of secondary structures
is not driven solely by the multiplicity of possible structures.
In fact, it is known that the secondary structures
obtained by energy minimizations 
tend to have more basepairs per stem than 
that is expected from combinatorics \cite{HSS_1998}.
This is mainly because, from the energy point of view,
a stack of basepairs contributes to the stability of the stem
in such a way that a longer stem is preferred.

In order to take account of the stack stability
in combinatorial approaches,
it is usual to consider so-called
$r$-canonical structures,
which are
the structures without a stem of length less than $r$ \cite{BLR_2016}.
Assuming such short stems are energetically unstable,
one may exclude the structures with the short stems
and investigate the space of only $r$-canonical structures
from the outset.
The $r$-canonical island diagrams 
can be built up from the single-stack diagrams
by assigning $r-1$ more basepairs at each stem
from the beginning:  
\be
F(u\,z^{r-1},x,y,z)=G\Big(x,y,\frac{u\, z^r}{1+ u\, z^r -z}\Big) \,.
\ee

\begin{figure}[tpb]
\centering
\includegraphics[width=0.46\textwidth]{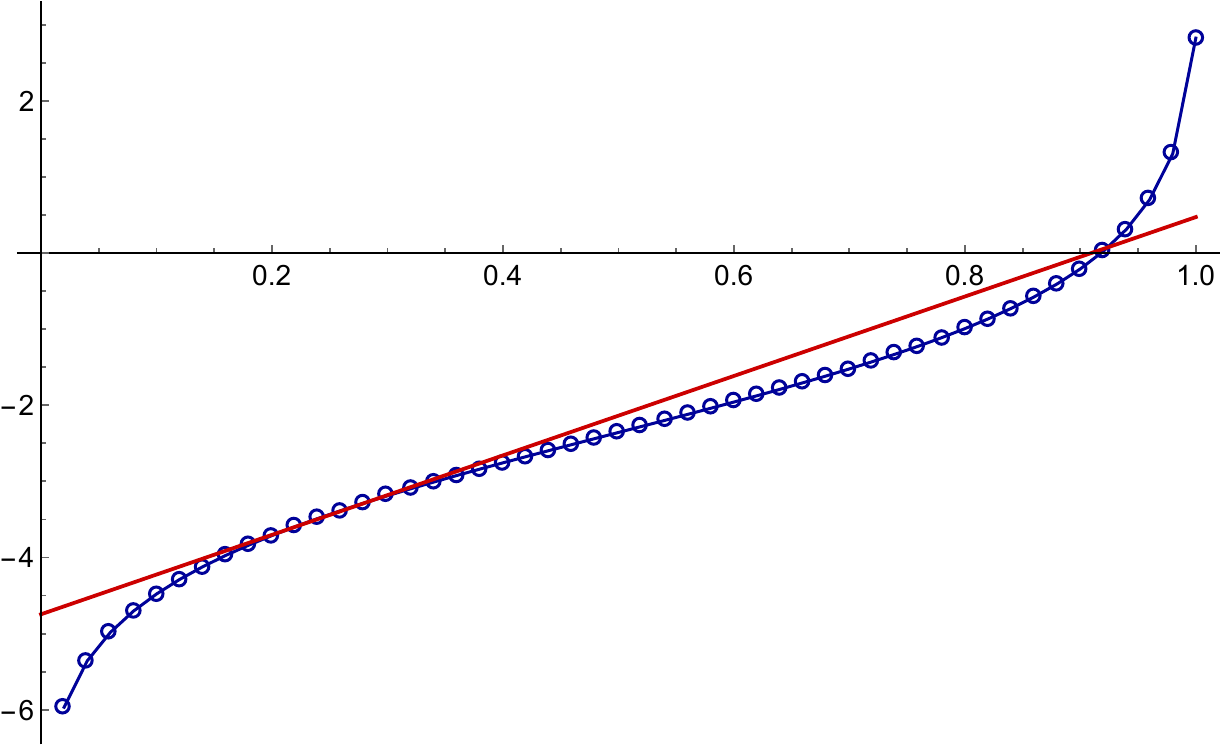}
\caption{\label{fig:mu} Chemical potential $\mu$ as a function of the peak
 value
$k_0/\ell$
for $\ell=100$. The linear fitting is shown for the range from 1/5 to 1/3.}
\end{figure}

On the other hand, however, one may consider another way
that allows short stems,
but with a certain weight reflecting their instability.
The weight can be imposed on the number of stems $k$
since the length of each stem decreases
as $k$ increases for given $\ell$.
We introduce $e^{\mu }$ as the weight
and put it in the generating function as
\be
F(u\, e^{\mu},x,y,z)
=G\Big(x,y,\frac{ u\, z \, e^{\mu}}{1+ u\, z \,e^{\mu}-z}\Big)
\ee
which is equivalent to replacing
$\Omega_{k,\ell}$ 
with $e^{\mu k}\Omega_{k,\ell} $.
One may find the system 
analogous to the grand canonical ensemble.
The weight is the fugacity
with $\mu$ playing the role of the chemical potential
related to creating stems.
Regarding $\log \Omega_{k,\ell} $ as entropy, 
one has $\mu = - \frac{\partial }{\partial k}\log \Omega_{k,\ell}$
at the maximum.
Depending on the value of $\mu$, one finds the 
peak value $k_0$ of $\Omega_{k,\ell} $  shifted 
(triangle or circle marks in Fig.\ref{fig:omega}).
The database \cite{rnadata}
shows that
RNA molecules 
tend to have the average stack length per stem in the range from 2 to 10,
mostly near $4$ independent of $\ell$. Referring to this,
one may require $k_0/\ell$ to be in the range from 1/5 to 1/3 and
finds $\mu$, linearly fitted as (Fig.\ref{fig:mu})
\be
\mu(k_0)\simeq -4.7+5.0\,  k_0/\ell
\ee
whose number is only slightly changed as $\ell$ varies.

\begin{figure}[!bt]
\centering
\includegraphics[width=0.48\textwidth]{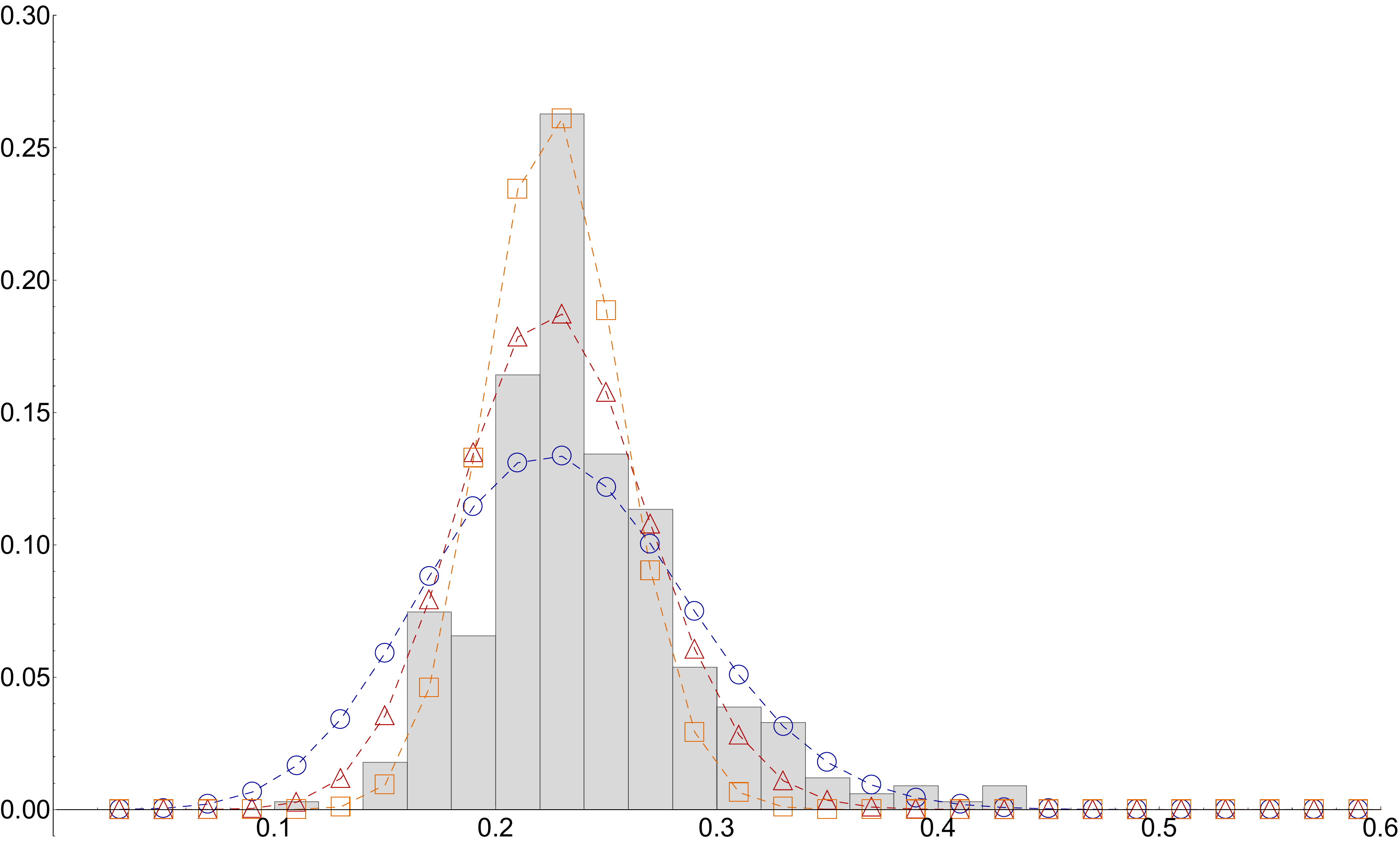}
\caption{\label{fig:data} 
Histogram of RNA for $\ell \geq 50$ obtained from \cite{rnadata}
as a function of $k/\ell$. Normalized distributions
$\Omega_{k,\ell}\, e^{\mu \, k}/(\sum_k \Omega_{k,\ell}\, e^{\mu \, k})$ 
for $\mu=-3.6$  when $\ell=50$(circle), $100$(triangle), $200$(square)
are given for comparison.}
\end{figure}

This idea is tested using the data
obtained from \cite{rnadata}.
The number of RNA molecules is given by a histogram as a function of $k/\ell$
where we set the interval of the bin to be 0.02 (Fig.\ref{fig:data})
and require $\ell \geq 50$ so that at each bin the data is not empty.
The total number of collected data turns out to be 335 from \cite{rnadata}
and they are distributed
as 283 ($50 \leq \ell < 150$) and 52 ($150 \leq \ell $).
The best sample will be the data set with given $\ell$.
However, we cannot get a significant number of data set with a given $\ell$.
To overcome the statistical error, we use the whole data set with $\ell \geq 50$
assuming different $\ell$ of the data set does not affect much of the result.
The histogram shows 
$0.22 \leq k_0/\ell < 0.24$ and therefore,  $\mu \simeq -3.6$. 
For comparison, the normalized distributions corresponding to $\mu=-3.6$
are plotted using the generating function 
when $\ell=50$(circle), $100$(triangle), $200$(square).
Different $\ell$ shows only the slight change of the standard deviation 
which confirms the assumption.

One way of interpreting the chemical potential
from the perspective of usual energy model 
is the stacking energy of base pairs.
It is believed that the attractive energy
between two consecutive base pairs
is the main contribution to the stability of the stem. 
The stacking energy depends not only on
the nucleotides involved but also on the stacking order of the two pairs along the backbone \cite{FKJSCNT_1986}.
Nevertheless, if we assume an average stacking energy $\epsilon$ at some fixed temperature,
we may have additional statistical factor $e^{\epsilon(\ell-k)}$ 
to $\Omega_{k,\ell}$ for the given structures with $k$ stems and $\ell$ basepairs.
When we consider the distribution for fixed $\ell$,
the factor $e^{\epsilon \ell}$ cancels out by the normalization.
Therefore, one may view the chemical potential
as the average stacking energy, $\mu=-\epsilon$.
We remark that
the number of ways putting unpaired nucleotides on bridges
can significantly change the value of $\epsilon$.
The value $-3.6$ of $\mu$ 
should be considered 
as the one evaluated under
the reference point 
that an appropriate number of unpaired nucleotides 
is already assigned on the bridges.

Another way is to understand
the stack stability 
in terms of the multiplicity of structures 
from the pure combinatorial point of view.
Recall that
an island diagram is in general
the abstract structure
of numerous secondary structures.
In this respect,
one may interpret
the factor $e^{\mu k}$
as the one reflecting the number
of secondary structures
which reduce to the island diagrams for a given $k$.
In other words,
the fugacity quantifies
how fast the multiplicity of secondary structures decreases
as $k$ increases.

%====================
\section{Possible extensions} 
\label{sec:pe}
%====================

In this section, we remark on some extensions
and possible applications
by presenting brief descriptions
without giving much detail.
Firstly, we mention that the relation
between island diagrams and
RNA abstract shapes,
which are provided in \cite{GVR_2004}
to classify secondary structures according
to their structural similarities.
In particular, the single-stack diagram is directly related to 
$\pi'$-shape.
Given a secondary structure, its $\pi'$-shape
is obtained as follows: each stem is replaced by a single basepair.
And each group of maximally consecutive unpaired bases
is represented as a single unpaired region,
regardless of the number of unpaired bases in it
(refer to \cite{GVR_2004, LPC_2008} 
for its formal definition and details).
The unpaired regions corresponds to the bridges
so that the single-stack diagrams are indeed the $\pi'$-shapes.
The only difference is that, in island diagrams,
we identify the bridge before the first island
with the one after the last island just for convenience.
Therefore, the generating function of single-stack diagrams
can immediately be applied to $\pi'$-shapes and one obtains
$(1+y)^2 S(x, y, z)/y$.
Here, $x$ and $z$ are the expansion variables for
the number of hairpins and basepairs(stems), respectively,
whereas $y$ now stands for the expansion variable of bridges.
The factor $(1+y)^2$ reflects the possible four cases
of putting bridges at the two ends of the backbone.

Although we have dealt with abstract structures
ignoring the number of unpaired bases
to analyze statistics for given $\ell$,
it is easy to derive the generating function
with the total number of bases
from the island diagrams
by simply putting the unpaired bases in bridges.
Let us introduce one more expansion variable $w$ 
corresponding to the number of bases.
The procedure is as follows:
first, the number $I-1-h$ of bridges 
between islands (excluding hairpin loops
and two bridges at the end of backbone)
must contain at least one nucleotide,
which will give  the factor $w^{I-1-h}$.
Second, each basepair has two nucleotides 
that implies the factor  $w^{2\ell}$
for given $\ell$. 
Third, there are $I+1$ bridges on the backbone
and one may put arbitrary number of nucleotides on each bridge,
which introduces the factor $1/(1-w)^{I+1}$.
In addition to this, one may introduce the constraint
that each hairpin loop consists of
at least $\lambda-1$ nucleotides,
which reflects the rigidity of the backbone.
This constraint shifts the variable $x$ to $x w^{\lambda-1}$.
As a result,
the generating function 
of a given number of nucleotides
can be obtained 
from the previous generating functions
by replacing the variables $x \to xw^{\lambda-2}$, $y \to yw/(1-w)$, $z \to zw^2$
and multiplying the overall factor $1/(w(1-w))$.
For instance, one obtains 
$r$-canonical generating function 
$R_\lambda(u\,z^{r-1},x,y,z;w)
:=\sum_{k, h,I,\ell,n} r_\lambda(k,h,I,\ell;n) \, u^k x^h y^I z^{\ell}w^n $,
\begin{align}
R_\lambda&(u\,z^{r-1},x,y,z;w) \\
 &=\frac1{w(1-w)}
 F\bigg(u(zw^2)^{r-1},x w^{\lambda-2},\frac{y\,w}{1-w},zw^2 \bigg) \,.  \nn
\end{align}
One can check that the generating function $R_\lambda(z^{r-1},1,1,z;w)$ 
agrees
with the one $S_\lambda^{[r]}(w,z)$ appeared in \cite{BLR_2016}
where stems, hairpin loops and  islands are not the observables.

\begin{comment}
?????????????????????
an RNA sequence is compatible with a structure s if it can in principle form this structure irrespective of energetic constraints

Bisecondary structures

intuitively obtained by drawing a one secondary structure in
the upper half-plane and another in the lower half plane such that each vertex
has degree at most 1. The bisecondary structure is then derived by “flipping”
the arcs contained the lower half plane “up”.

no generating function is known.

a classification of their knot types based on a notion
of inconsistency graphs and gave an upper bound for bisecondary structures.
What constitutes the main difficulty here is the lack of an inductive recurrence
relation,

???????????????????????????????
\end{comment}

\begin{figure}[!h]
\vspace*{0.5cm}
\centering
\includegraphics[width=0.42\textwidth]{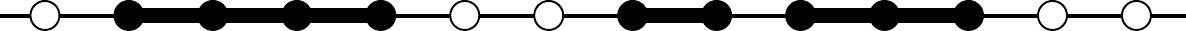}
\caption{\label{fig:lattice}
A particular configuration in a polydisperse system
that consists of three chains with 4, 2 and 3 monomers
on the lattice with 14 sites.
The empty circles represent empty sites.
}
\end{figure}

Finally, we want to remark here on an
interesting possible application.
The configuration of island diagrams
is seen to have 
a close relation to
the model of polydisperse chains on a one-dimensional lattice
\cite{JMW_2006}.
In the model, 
a chain with the number $M$ of monomers is placed on a lattice
in such a way that one monomer of the chain occupies one lattice point.
The polydisperse system consists of
numerous chains having diverse sizes
placed on a lattice (Fig.\ref{fig:lattice}). 
In \cite{JMW_2006},
the entropy of the system
and the distribution of chain sizes
are calculated as a function of the density of monomers
when the lattice length is taken to be large.

One can immediately make an analogy between
the polydisperse system and the island diagrams.
When only the sequence of islands without base pairs 
is taken into account,
an island of length $M$ 
corresponds to a chain with $M$ monomers
and the number of islands
matches up with the number of chains.
The number of nucleotides is the lattice length
and unpaired ones are empty sites on the lattice
such that the density of monomers 
is the density of paired ones.
Having the analogy in mind, let us briefly 
illustrate the generating functions 
for the polydisperse system.
Let $\mathfrak{g}(I,L)$ denote the number of 
configurations with $I$ chains(islands) and $L$ monomers(paired bases).
Since there are no concepts of hairpins or base pairs,
$\mathfrak{g}(I,L)$ is just given by the summation of $1$ over 
$k_1+\cdots+k_I=L$
such that its generating function is given as
\be
\mathfrak{G}(y,z):=\sum_{I,L \geq 0} \, \sum_{k_1+\cdots+k_I=L} y^I \, z^L 
=\frac{1}{1-\frac{y}{1-z}} \,.
\ee
Now we take empty sites(unpaired bases) into account
by putting them between chains.
The polydisperse system has no restrictions
on the minimum number of empty sites
between chains
and hence it is simpler than the case of island diagrams given above.
Referring to the above procedure,
one may easily find the generating function,
\be
\mathfrak{R}(y,z;w):=\sum_{I,L,N} \mathfrak{r}(I,L;N) y^I z^L w^N
=\frac{1}{1-w-\frac{y z w}{1-z w}} \,.
\ee
where $\mathfrak{r}(I,L;N)$ denotes
the number of configurations with
$I$ chains, $L$ monomers and $N$ lattice length(total number of nucleotides).

One may exploit the generating function to
find asymptotic behaviors of the coefficients
when the lattice length is taken to infinity
and reproduce known results given in \cite{JMW_2006}
such as the distribution of chain sizes.
Furthermore, it will be an amusing study
to consider the role of base pairs
in connection with the system of chains.
In particular, 
the concept of hairpins
may have an immediate application
as is related to
an interaction between adjacent chains.

%===================
\section{Conclusion} 
\label{sec:con}
%====================
We considered 
the island diagrams which are
the abstraction of RNA secondary structures.
The island diagram is introduced to 
study combinatorics of the secondary structures
for a given number of base pairs,
rather than for a given number of nucleotides.
Using the Hermitian matrix model
with the help of Chebyshev polynomial,
we can derive various generating functions 
of the structures in a closed form.
In addition, we introduced 
the fugacity to match the experimental finding 
of average number of basepairs in a stem. 
As a result, we evaluated the chemical potential of the stem
from the combinatorial approach.
Finally, we also suggested possible extensions
of the island diagrams.
In particular, 
the application 
to the model of polydisperse chains will be of interest.

\begin{acknowledgments}
The authors acknowledge the support of this work by the National Research
Foundation of Korea(NRF) grant funded by the Korea government(MSIP) (NRF-2017R1A2A2A05001164).
S.K. Choi is partially supported by National Science Foundation of China
(NSFC) under the project 11575119.
\end{acknowledgments}

\end{document}